\documentstyle[12pt]{article}
\begin{document}
\baselineskip=20pt
\begin{center}
{\Large \bf Chiral Symmetry Breaking \\
in the Dual Ginzburg-Landau Theory}
\end{center}
\begin{center}
{\rm
Hiroshi Toki, Shoichi Sasaki, Hiroko Ichie and Hideo Suganuma\\}
{\it  
Research Center for Nuclear Physics (RCNP), Osaka University\\
Mihogaoka 10-1, Ibaraki, Osaka 567, Japan
}
\end{center}
\begin{quote}
\baselineskip=15pt
{\small
Confinement and chiral symmetry breaking are the most fundamental
phenomena in Quark Nuclear Physics, where hadrons and nuclei are 
described in terms of quarks and gluons.  The dual Ginzburg-Landau
(DGL) theory, which contains monopole fields as the most essential 
degrees of freedom and their condensation in the vacuum, is modeled 
to describe quark confinement in strong connection with QCD.  
We then demonstrate that the DGL theory is able to describe the
spontaneous break down of the chiral symmetry.
} 
\end{quote}
\baselineskip=20pt
\section{Introduction}

In Quark Nuclear Physics (QNP), the most essential phenomena are confinement 
of quarks and gluons and the chiral symmetry breaking. 
Quarks are not found in free space,
while they are seen in deep inelastic scattering. 
Quarks are present in hadrons and hence the understanding of confinement 
is essential for the sizes of hadrons.  
	 
The chiral symmetry is found in the QCD lagrangian. 
In the u-d sector, the current mass is considered negligible ($\sim$5MeV) 
as compared to the hadron mass ($\sim$1GeV), and hence the chiral symmetry
is to be realized with high accuracy.  We expect then the chiral 
partners, saying simply parity doublets, to be degenerate in the u-d meson 
spectrum.
The nature, on the other hand, shows that the pion ($0^-$) has 139MeV 
and no $0^+$ partner is found.
The rho meson ($1^-$) is 500MeV apart from its
chiral partner, the a1 meson ($1^+$).
The same feature is found also for the baryon spectrum. 
This property of hadron spectra tells us that the chiral symmetry
is broken spontaneously and the pion appears 
as its Goldstone particle.
This chiral symmetry breaking should be
understood for full description of QNP. 
Chiral symmetry breaking  would provide the constituent quark masses
and even the small mass pions, which are responsible for N-N interaction.

How do these phenomena happen? 
For this let us see the QCD coupling strength $\alpha_S$,
which runs with relevant 
momentum scale obeying the renormalization group of QCD\cite{1}.
At large momenta, $\alpha_S$  diminishes and the theory becomes asymptotic free, 
and hence the momentum dependence of deep inelastic scattering 
is calculable perturbatively. 
At small momenta, where confinement and chiral symmetry breaking are 
expected, $\alpha_S$  blows up and hence we face highly non-perturbative
processes.  
We ought to find the essential degrees of freedom to get the insight of 
these phenomena.

\section{Dual Ginzburg-Landau theory}

It is Nambu, who introduced an interesting view on color confinement in 
1974\cite{2}.
Suppose we insert a superconductor in a magnetic field.  
The superconductor does not allow the magnetic field to pass through. 
If it were to allow the magnetic field, in the superconductor of second kind, 
the magnetic field should be confined in a vortex-like configuration.  
This is known as the Meissner effect.  Nambu takes its dual version for 
quark confinement. If the vacuum is normal, the color electric field 
should look like the one of the Coulomb potential between a positive and a 
negative color charges.  If the vacuum is superconductor-like (dual 
superconductor), then the vacuum dislikes the electric field to pass 
through and hence the color electric flux ought to be confined in a 
vortex-like configuration.  This is then named as dual Meissner effect.  
This picture, however, does not become popular, because it requires color 
magnetic monopoles.  In the superconductor, the charged object, Cooper 
pair, is to condense, while in the QCD vacuum, the magnetically charged 
object, color magnetic monopole, to condense.

't Hooft was the one, who demonstrated the natural appearance of color 
magnetic monopoles in QCD\cite{3}.
In the non-abelian gauge theory like QCD, he 
introduced a particular gauge named abelian gauge, to reduce it to the 
abelian gauge theory like QED.  From a topological argument, color 
magnetic monopoles appear naturally in the abelian space.  This work then 
supports the idea  of Nambu for confinement.  Hence, QCD naturally reduces 
to QED with magnetic monopoles, which is the Maxwell equation with 
magnetic charges and currents, studied by Dirac\cite{4}.
This Maxwell equation has the duality symmetry,
which naturally arises in the special gauge of QCD.

	It took then about 10 years before the above idea was formulated in the 
form of lagrangian\cite{5}.
The dual Ginzburg-Landau (DGL) theory is expressed with the following
 lagrangian,
\begin{eqnarray}
{\cal L}_{DGL}={\cal L}_{\rm dual}
+\bar q(i\gamma _\mu \partial ^\mu -m+e\gamma _\mu A^\mu )q+
{\rm tr}[\hat D_\mu ,\chi ]^{\dag}[\hat D^\mu ,\chi ]-
\lambda {\rm tr}(\chi^{\dag} \chi -v^2)^2
\end{eqnarray}
Here, ${\cal L}_{\rm dual}$  denotes the dual version of the gauge field tensor. 
$q$ is the quark field and $\chi$  is the monopole field.
$\hat D^\mu$  is the dual covariant derivative,  $\hat D^\mu=\partial^\mu 
+ ig B^\mu$.
$B^\mu$  is the dual gauge field and $g$ is the dual coupling 
constant, which satisfies the Dirac condition, $eg=4\pi$.
The last term is the 
Higgs term to cause monopole condensation, where $\lambda$ and $v$  are the 
parameters of the DGL lagrangian.  It is important to note that this DGL 
lagrangian is derived from the QCD lagrangian by assuming the existence of 
the monopole field and the abelian dominance\cite{5}.  It is at the same time 
supported by the recent lattice QCD calculations\cite{6}.
This is the dream 
lagrangian of Dirac, which ought to appear in some non-abelian gauge 
theory like QCD.
  
	The first application is the $q \bar q$ static potential by putting
$q \bar q$	pair with distance $r$\cite{7}.
The potential comes out to have a Yukawa term and a linear 
confining term. We can fix the parameters of the DGL lagrangian by fitting 
to the phenomenological potential.  As for the glueball mass, which 
appears in the DGL theory and has a strong connection with the QCD vacuum 
and confinement, it turns out M($0^+$) $\sim$ 1.5 GeV.
Its appearance of the 
linear potential is not surprising, since it is modeled in the DGL 
theory.  It is worthwhile to stress, however, that there are no other 
models, which are able to realize confinement of colors and at the same 
time have a strong link with QCD.  The real challenge is now the chiral 
symmetry breaking, which is discussed next.

\section{Chiral symmetry breaking}

	 Chiral symmetry breaking is directly related with the quark mass 
generation in the QCD vacuum.  How quarks then behave in monopole 
condensed vacuum?  It corresponds to solving the Schwinger-Dyson equation, 
where quarks get the self-energy corrections due to the non-perturbative 
interaction with gluons\cite{7}.
This seems, however, unphysical, because quarks 
are confined.  It means that whenever a quark is present, there should be 
an anti-quark or a di-quark to make the system color-singlet.  In 
principle, therefore, we ought to solve multi-body system to talk about a 
single quark.  Suppose we have the Schwinger-Dyson equation as 
schematically written as,
\begin{eqnarray}
S^{-1}(p)=S_0^{-1}(p)+
\int_0^\infty  S(p-q)D(q)dq.
\end{eqnarray}
The quark, which is confined, should be seen from the position of 
anti-quark.  Then, the probability to find the quark is finite only within 
the distance of the hadronic scale; $\sim$1fm.  Therefore, gluons cannot travel 
freely any distance.  Rather, it is confined also within the hadronic 
distance.  This indicates that there should be the infrared cut-off, which 
is of order of the inverse of the confining distance $R$ as
$q > q_c =\frac1R$.  Hence, the 
SD equation is modified simply to 
\begin{eqnarray}
S^{-1}(p)=S_0^{-1}(p)+
\int_{q_c}^\infty  {S(p-q)D(q)dq}.
\end{eqnarray}
We show in Fig.1 the result of chiral symmetry breaking, expressed in 
terms of the quark mass $M(p)$.  It becomes finite with increasing the strength 
of monopole condensation.  We find also the pion decay constant and the 
quark condensate to have the numbers close to the semi-experimental 
values.  This calculation demonstrates that monopole condensation is the 
source of both the confinement and the chiral symmetry breaking\cite{8}.

We discuss also the recovery of the chiral symmetry at finite 
temperature\cite{9}.
We can formulate this in the imaginary time formalism.  We 
in fact find the recovery of the chiral symmetry as indicated in Fig.2 by 
the ratio of quark condensate at finite and zero temperatures 
$\langle \bar q q\rangle_T$. $\langle \bar q q\rangle_T$    
decreases with temperature and eventually drops to zero, indicating the 
recovery of the chiral symmetry.  We note that the temperature of the 
phase transition; $T_c \sim$0.11GeV
seems smaller than the results of lattice 
QCD.  This difference would be caused by the use of temperature 
independent parameters in the Higgs term.  Since this term is introduced 
at zero temperature, it is likely that they depend on temperature as the 
case of the superconductor.  In addition, the hadronic scale should also 
depend on temperature.  Here, the point of showing this result is merely 
to demonstrate that the DGL theory provides phase transition to the normal 
phase at finite temperature.

	We can even talk about the confinement-deconfinement phase transition at 
finite temperature\cite{10}.  In the quenched approximation, we can write the DGL 
lagrangian in terms of the dual gauge fields by integrating out the gauge 
fields; $A_\mu$.  It amounts to calculate the partition functional and we can 
derive the effective potential; the therodynamical potential.  The result 
is shown in Fig.3, where the effective potential is plotted as a function 
of the monopole condensate, $\chi$.  The absolute minimum appears at finite 
value of  $\chi$ in the lower temperature side and jumps to   $\chi$=0.  This indicates 
deconfinement phase transition of first order.  We can calculate the 
string tension as a function of temperature, the result of which is shown 
in Fig. 4.  Adjusting $\lambda$ as a temperature dependent parameter so as to 
reproduce the critical temperature at 0.2GeV, we find the string tension 
to reproduce the lattice QCD results\cite{11}.

	It is interesting to mention that lattice QCD is also doing a very 
interesting development.  The $q \bar q$  potential calculated 
with full lattice QCD 
agrees with the result of only the abelian part.  This agreement indicates 
that the confinement physics is describable in terms of only the abelian 
gluons.  Other quantities in this direction are being done by several 
groups\cite{12,13}.  All these results indicate that the low momentum phenomena 
could be described by the abelian gauge gluons (abelian dominance), when 
the abelian gauge is suitably chosen.  At RCNP, we have started numerical 
experimental program with the use of lattice QCD\cite{14}.
The largest task is 
to predict the properties of the glueballs\cite{15} associated with confinement, 
particularly the decay properties for experimental identification.

\section{Conclusion}

	Quark Nuclear Physics (QNP) is the field to describe nucleons, mesons and 
nuclei in terms of quarks and gluons.  The most essential phenomena in QNP 
are confinement of quarks and gluons and the chiral symmetry breaking.  
The confinement is modeled as due to the dual Meissner effect and is 
expressed in terms of the dual Ginzburg-Landau theory, where the QCD 
monopoles and their condensation in the QCD vacuum are the essential 
ingredients.  We have demonstrated in this paper that the DGL theory is 
able also to describe the chiral symmetry breaking.  We have then 
discussed the recovery of these symmetries at finite temperature.  Now the 
DGL theory is ready for exciting experimental phenomena of QNP.

	The author is grateful to the members of the theory group of RCNP for 
exciting collaborations of the theoretical accounts of the Quark Nuclear 
Physics.  We mention also the strong support of the director of RCNP, 
Prof. H. Ejiri, for continuous encouragement of the theoretical works.  
H.T appreciates the fruitful organization of this Workshop and thanks A. 
Thomas and A. Williams for the nice organization and their hospitalities 
to all the Japanese participants.

	This is an invited talk presented by H. Toki in the Joint Japan-Australia 
Workshop, held in Nov. 15-24, 1995 organized by Adelaide Institute of 
Theoretical Physics.

\newpage

{\bf \large Figure Captions}
\vspace{0.3cm}\\
Fig.1 The constituent quark mass calculated within the DGL theory with 
various values of the dual gauge mass, which indicates the strength of 
monopole condensation, as a function of the Euclidean momentum square.  
The unit $\Lambda_{\rm QCD}$ is 200 MeV.\\
Fig.2 The ratio of quark condensate at finite and zero temperatures within 
the DGL theory as a function of temperature.  The critical temperature is 
about 0.11GeV.\\
Fig.3 The effective potential (thermodaynamical potential) at various 
temperatures within the DGL theory as a function of the monopole 
condensate.  The absolute minimum is indicated by  $\times$ for each curve, 
indicating the jump (phase transition of first order) around $T =$ 0.5GeV.\\
Fig.4 The string tension between a quark and antiquark pair for constant 
and variable lambda's within the DGL theory as a function of temperature.  
The dots are the results of the pure-gauge lattice QCD; \cite{10}.

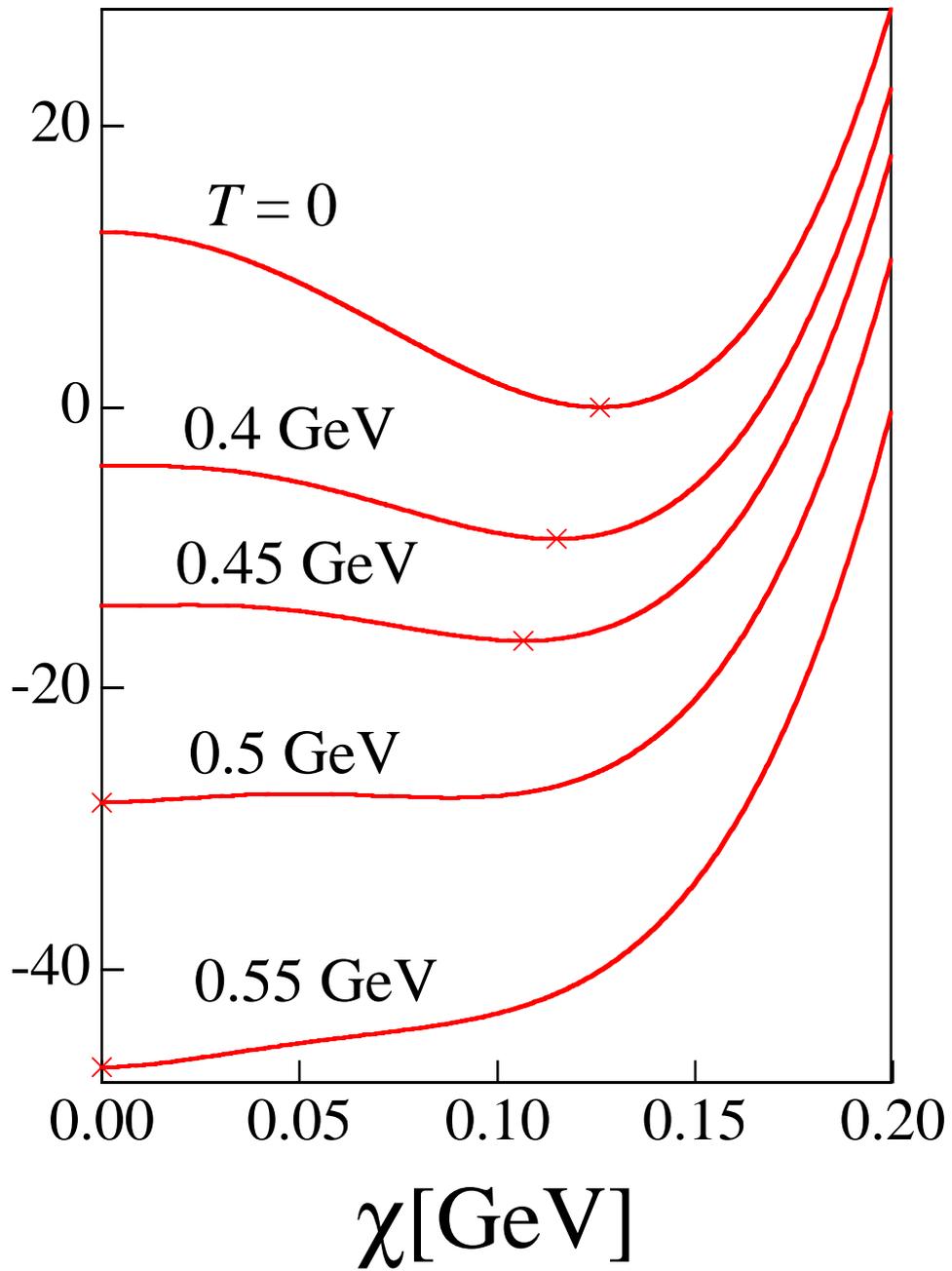
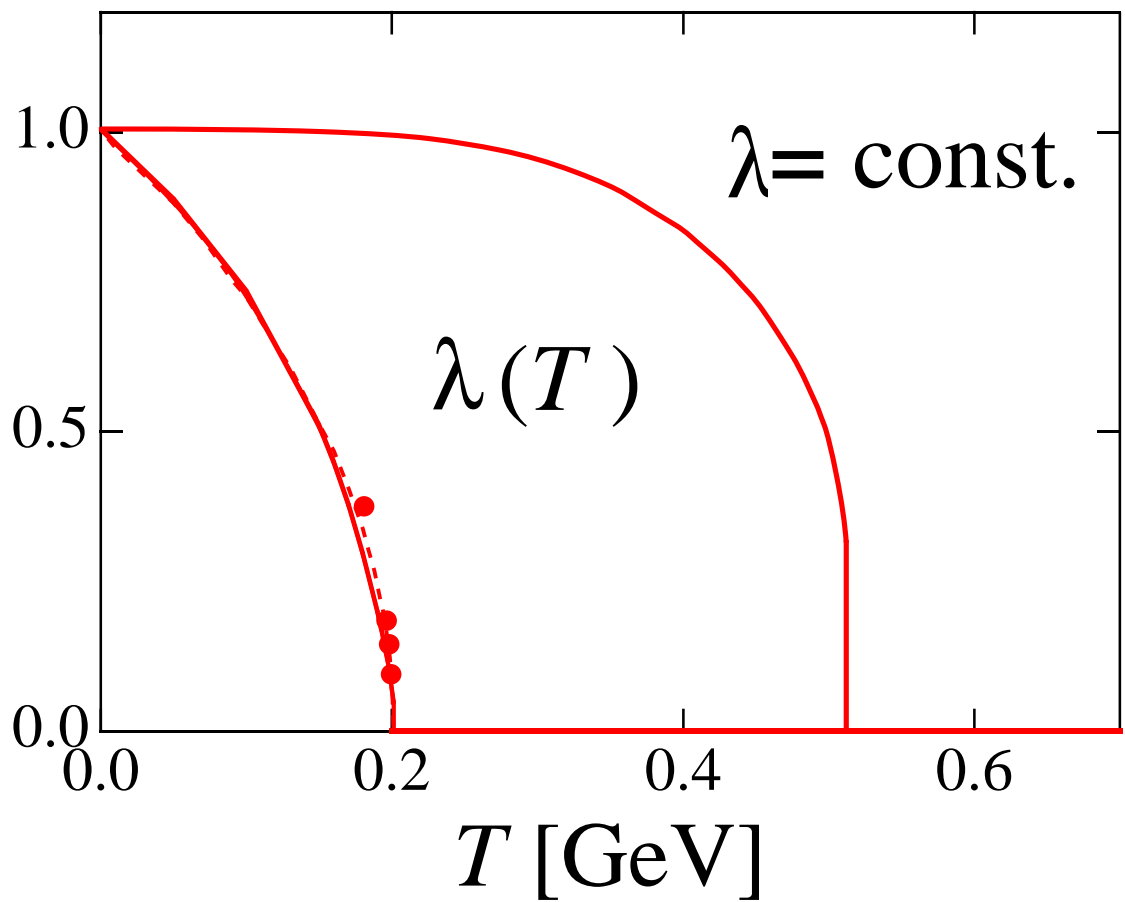
\end{document}